\documentclass[twocolumn, times]{aastex631}
\usepackage{xspace}
\usepackage{amsmath}
\usepackage{amssymb}
\usepackage{bm}
\usepackage{hyperref}

\newcommand{\tess}{\emph{TESS}\xspace}

\providecommand{\bjdtdb}{\ensuremath{\rm {BJD_{TDB}}}}

\providecommand{\feh}{\ensuremath{\left[{\rm Fe}/{\rm H}\right]}}

\providecommand{\rj}{\ensuremath{R_{\rm Jup}}}
\providecommand{\mj}{\ensuremath{M_{\rm Jup}}}

\newcommand{\ms}{\,m\,s$^{-1}$}

\newcommand{\tic}{TIC 324609409\xspace}

\newcommand{\toi}{TOI-6692\xspace}
\newcommand{\toib}{TOI-6692 b\xspace}

\newcommand{\pmRA}{38.7968\pm0.0128}
\newcommand{\pmDEC}{-22.7508\pm0.0128}
\newcommand{\parallax}{3.2203\pm0.0119}

\newcommand{\loggfit}{4.176^{+0.054}_{-0.051}}
\newcommand{\tefffit}{5890^{+170}_{-180}}
\newcommand{\metfit}{-0.008^{+0.030}_{-0.027}}
\newcommand{\mstar}{1.047^{+0.089}_{-0.071}}
\newcommand{\rstar}{1.385^{+0.054}_{-0.052}}
\newcommand{\lstar}{2.08\pm0.20}
\newcommand{\age}{7.8^{+3.1}_{-3.0}}
\newcommand{\dist}{310.5^{+1.2}_{-1.1}}

\newcommand{\per}{130.57^{+0.42}_{-0.35}}
\newcommand{\plrad}{1.042^{+0.050}_{-0.049}}
\newcommand{\plmass}{0.620^{+0.080}_{-0.065}}
\newcommand{\ecc}{0.537\pm0.061}
\newcommand{\semimaj}{0.512^{+0.014}_{-0.012}}

\providecommand{\bjdtdb}{\ensuremath{\rm {BJD_{TDB}}}}

\providecommand{\feh}{\ensuremath{\left[{\rm Fe}/{\rm H}\right]}}

\providecommand{\mj}{\ensuremath{\,M_{\rm J}}}
\providecommand{\rj}{\ensuremath{\,R_{\rm J}}}
\providecommand{\me}{\ensuremath{\,M_{\rm E}}}

\accepted{January 22, 2026}

\shorttitle{TOI-6692: long period single-transit giant planet}
\shortauthors{Bieryla, et al.}

\begin{document}

\title{TOI-6692b: An eccentric 130 day period giant planet with a single transit from \emph{TESS}}

\correspondingauthor{Allyson Bieryla}
\email{abieryla@cfa.harvard.edu}

\newcommand{\CfA}{Center for Astrophysics \textbar \ Harvard \& Smithsonian, 60 Garden Street, Cambridge, MA 02138, USA}
\newcommand{\USQ}{University of Southern Queensland, Centre for Astrophysics, West Street, Toowoomba, QLD 4350 Australia}
\newcommand{\cs}{Citizen scientist, c/o Zooniverse, Department of Physics, University of Oxford, Denys Wilkinson Building, Keble Road, Oxford, OX1 3RH, UK}
\newcommand{\CarnegieOBS}{The Observatories of the Carnegie Instution for Science, 813 Santa Barbara Street, Pasadena, CA, 91101, USA}
\newcommand{\CarnegieEPL}{Earth and Planets Laboratory, Carnegie Institution for Science, 5241 Broad Branch Road, NW, Washington, DC 20015, USA}
\newcommand{\MITKavli}{Department of Physics and Kavli Institute for Astrophysics and Space Research, Massachusetts Institute of Technology, Cambridge, MA 02139, USA}

\author[0000-0001-6637-5401]{Allyson Bieryla}
\affiliation{\CfA}
\affiliation{\USQ}

\author[0000-0001-6588-9574]{Karen A.\ Collins}  
\affil{\CfA} 

\author[0000-0002-4891-3517]{George Zhou} 
\affiliation{\USQ}

\author[0000-0001-9911-7388]{David W. Latham} 
\affil{\CfA}

\author[0000-0003-0035-8769]{Brad Carter} 
\affil{\USQ}

\author[0000-0002-4297-5506]{Paul Dalba} 
\affiliation{Department of Astronomy and Astrophysics, University of California, Santa Cruz, CA 95064, USA}

\author[0000-0002-5665-1879]{Robert Gagliano} 
\affiliation{Amateur Astronomer, Glendale, Arizona 85308}

\author[0000-0003-3988-3245]{Thomas L. Jacobs} 
\affiliation{Amateur Astronomer, Missouri City, Texas, 77459, USA}

\author{Martti Holst Kristiansen} 
\affiliation{Brorfelde Observatory, Observator Gyldenkernes Vej 7, DK-4340 Tølløse, Denmark}

\author{Daryll LaCourse} 
\affiliation{Amateur Astronomer}

\author{Mark Omohundro} 
\affil{\cs}

\author[0000-0002-1637-2189]{H.M. Schwengeler} 
\affil{\cs}




\author[0000-0003-1464-9276]{Khalid Barkaoui}
\affiliation{Instituto de Astrof\'isica de Canarias (IAC), Calle V\'ia L\'actea s/n, 38200, La Laguna, Tenerife, Spain}
\affiliation{Astrobiology Research Unit, Universit\'e de Li\`ege, 19C All\'ee du 6 Ao\^ut, 4000 Li\`ege, Belgium}
\affiliation{Department of Earth, Atmospheric and Planetary Science, Massachusetts Institute of Technology, 77 Massachusetts Avenue, Cambridge, MA 02139, USA}

\author[0000-0002-9158-7315]{Rafael Brahm} 
\affil{Facultad de Ingenier\'ia y Ciencias, Universidad Adolfo Ib\'{a}\~{n}ez, Av. Diagonal las Torres 2640, Pe\~{n}alol\'{e}n, Santiago, Chile}
\affil{Millennium Institute for Astrophysics, Nuncio Monse\~{n}or Sotero Sanz 100, Of. 104, Providencia, Santiago, Chile}

\author[0000-0003-1305-3761]{R. Paul Butler} 
\affil{\CarnegieEPL}

\author[0000-0003-1963-9616]{Douglas A. Caldwell}
\affiliation{SETI Institute, Mountain View, CA 94043, USA}
\affiliation{NASA Ames Research Center, Moffett Field, CA 94035, USA}

\author[0000-0002-5226-787X]{Jeffrey D. Crane} 
\affil{\CarnegieOBS}

\author[0000-0002-6939-9211]{Tansu Daylan}
\affiliation{Department of Physics and McDonnell Center for the Space Sciences, Washington University, St. Louis, MO 63130, USA}

\author[0009-0002-9833-0667]{Sarah Deveny} 
\affiliation{Bay Area Environmental Research Institute, Moffett Field, CA 94035, USA}

\author[0000-0003-3773-5142]{Jason D.\ Eastman} 
\affil{\CfA}

\author[0000-0002-2036-2311]{Yadira S. Gaibor}
\affiliation{Department of Physics, Massachusetts Institute of Technology, 77 Massachusetts Avenue, Cambridge, MA 02139, USA}
\affiliation{\MITKavli}

\author[0000-0003-1462-7739]{Micha\"el Gillon} 
\affiliation{Astrobiology Research Unit, Université de Liège, Allée du 6 Août 19C, B-4000 Liège, Belgium}

\author[0000-0002-1493-300X]{Thomas Henning} 
\affiliation{Max-Planck-Institut für Astronomie, Königstuhl 17, D-69117 Heidelberg, Germany}

\author[0000-0003-1728-0304]{Keith Horne}
\affiliation{SUPA Physics and Astronomy, University of St. Andrews, Fife, KY16 9SS Scotland, UK}

\author[0000-0002-2532-2853]{Steve B. Howell} 
\affiliation{NASA Ames Research Center, Moffett Field, CA 94035 USA}

\author[0000-0001-8923-488X]{Emmanuel Jehin} 
\affiliation{Space Sciences, Technologies and Astrophysics Research (STAR) Institute, Université de Liège, Allée du 6 Août 19C, B-4000 Liège, Belgium}

\author[0000-0002-4625-7333]{Eric L.\ N.\ Jensen}
\affiliation{Department of Physics \& Astronomy, Swarthmore College, Swarthmore PA 19081, USA}

\author[0000-0002-5389-3944]{Andr\'es Jord\'an}
\affil{Facultad de Ingenier\'ia y Ciencias, Universidad Adolfo Ib\'{a}\~{n}ez, Av. Diagonal las Torres 2640, Pe\~{n}alol\'{e}n, Santiago, Chile}
\affil{Millennium Institute for Astrophysics, Nuncio Monse\~{n}or Sotero Sanz 100, Of. 104, Providencia, Santiago, Chile}

\author[0000-0001-9269-8060]{Michelle Kunimoto}
\affiliation{Department of Physics and Astronomy, University of British Columbia, 6224 Agricultural Road, Vancouver, BC V6T 1Z1, Canada}

\author{Colin Littlefield} 
\affiliation{Bay Area Environmental Research Institute, Moffett Field, CA 94035, USA}


\author[0000-0002-7382-1913]{Léna Parc}
\affiliation{Observatoire de Genève, Département d’Astronomie, Université de Genève, Chemin Pegasi 51, 1290 Versoix, Switzerland}

\author[0000-0002-8964-8377]{Samuel N. Quinn} 
\affil{\CfA}

\author[0000-0002-7670-670X]{Malena Rice}
\affiliation{Department of Astronomy, Yale University, 219 Prospect Street, New Haven, CT 06511, USA}

\author[0000-0001-8812-0565]{Joseph E. Rodriguez} 
\affiliation{Center for Data Intensive and Time Domain Astronomy, Department of Physics and Astronomy, Michigan State University, East Lansing, MI 48824, USA}

\author[0000-0001-8227-1020]{Richard P. Schwarz}
\affiliation{\CfA}

\author[0000-0003-3904-6754]{Ramotholo Sefako} 
\affiliation{South African Astronomical Observatory, P.O. Box 9, Observatory, Cape Town 7935, South Africa}

\author{Stephen A. Shectman} 
\affil{\CarnegieOBS}

\author[0000-0002-1836-3120]{Avi Shporer}
\affiliation{\MITKavli}

\author[0000-0002-0345-2147]{Abderahmane Soubkiou} 
\affiliation{Astrobiology Research Unit, Université de Liège, Allée du 6 Août 19C, B-4000 Liège, Belgium}

\author{Gregor Srdoc}
\affil{Kotizarovci Observatory, Sarsoni 90, 51216 Viskovo, Croatia}

\author[0000-0003-3036-3585]{Michal Steiner} 
\affiliation{Observatoire de Genève, Département d’Astronomie, Université de Genève, Chemin Pegasi 51, 1290 Versoix, Switzerland}

\author[0009-0004-8891-4057]{Marcelo Tala Pinto} 
\affil{Department of Astronomy, McPherson Laboratory, The Ohio State University, 140 W 18th Ave, Columbus, Ohio 43210, USA}

\author{Johanna Teske} 
\affil{\CarnegieEPL}

\author[0000-0002-0236-775X]{Trifon Trifonov} 
\affiliation{Max-Planck-Institut für Astronomie, Königstuhl 17, D-69117 Heidelberg, Germany}
\affiliation{Department of Astronomy, Sofia University ``St Kliment Ohridski'', 5 James Bourchier Blvd, BG-1164 Sofia, Bulgaria}
\affiliation{Landessternwarte, Zentrum f\"ur Astronomie der Universit\"at Heidelberg, K\"onigstuhl 12, D-69117 Heidelberg, Germany}

\author{Solène Ulmer-Moll} 
\affiliation{Leiden Observatory, Leiden University, P.O. Box 9513, 2300 RA, Leiden, The Netherlands}
\affiliation{Observatoire de Genève, Département d’Astronomie, Université de Genève, Chemin Pegasi 51, 1290 Versoix, Switzerland}

\author[0000-0001-8621-6731]{Cristilyn N.\ Watkins}
\affiliation{Bozeman, MT 59718, USA}

\author[0000-0002-6937-9034]{Sharon X.~Wang}
\affiliation{Department of Astronomy, Tsinghua University, Beijing 100084, People's Republic of China}

\author[0000-0001-9261-8366]{Jhon Yana Galarza}
\altaffiliation{Carnegie Fellow}
\affiliation{The Observatories of the Carnegie Institution for Science, 813 Santa Barbara Street, Pasadena, CA 91101, USA}
\affiliation{Departamento de Astronomía, Facultad de Ciencias Físicas y Matemáticas Universidad de Concepción, Av. Esteban Iturra s/n Barrio Universitario, Casilla 160-C, Chile}

\author[0000-0001-7961-3907]{Samuel W. Yee} 
\affil{\CfA}

\begin{abstract}

We report the discovery and characterization of \toib, an eccentric ($e\sim0.54$) Jupiter on a 130-day orbit. \toib was first detected as a community TESS Object of Interest (cTOI) by the Visual Survey Group and the Planet Hunters group as a single transit candidate via TESS observation. The period was subsequently confirmed via radial velocity monitoring from the Planet Finder Spectrograph on the 6.5\,m Magellan telescope. Additional radial velocities were acquired with the CHIRON, FEROS, and CORALIE spectrographs. LCOGT ground-based photometric follow-up was conducted over 2 weeks to detect another transit and refine the period. Although we did not detect an ingress or egress of the 11.04 hr transit, we did detect a possible in-transit signal in the multi-night data and provide an updated ephemeris for future monitoring. \toib{} is one of few planets with orbital periods longer than 100 days that have a secure mass, radius, and eccentricity detection. As with most giant planets at these orbital periods, the eccentricity of \toib{} is lower than that expected of planets undergoing high-eccentricity tidal migration, but is more consistent with the expectations of planet-planet scattering outcomes. A long-term radial velocity trend was detected and further monitoring is warranted to determine the outer companion period. \toib{} is also one of few \emph{TESS} single transit targets that have its period eventually confirmed via follow-up photometric campaigns timed to capture transits despite the relatively large ephemeris uncertainties. Such efforts highlight the capabilities of night-to-night stability on ground-based photometric facilities today. 

\end{abstract}



\section{Introduction} \label{sec:Intro}

Less than 1\% of known giant exoplanets with periods longer than 100 days have their mass and radius measured\footnote{NASA Exoplanet Archive as of October 20, 2025}. Long-period exoplanets are inherently difficult to confirm. Observationally, transits of long-period planets are geometrically rare, and they require more dedicated time to follow up as a result of the rarity and long duration of their transits. As such, only 15 giant planets with orbital periods $>100$\,days have $3\sigma$ masses and radii measured. While these cool Jupiter planets are observationally challenging, they are important to invest resources into. Many such long-period planet candidates from NASA's Transiting Exoplanet Survey Satellite (\textit{TESS}; \citealt{ricker2015}) exhibit only single transit detections in the original discovery light curves. Following up such single-transit candidates is a technically challenging and resource intensive exercise. Of the 15 giant planets (listed in Table\,\ref{tab:longplanets}), five are single transit planet candidates that received campaigns of ground- and space-based follow-up to determine the true periods of the planets. 


Hot Jupiters experience high stellar irradiation, which can alter their thermal structure and size. Modeling this ``radius anomaly" effect is an active area of research \citep{baraffe2010, laughlin2018, 2021A&A...645A..79S}. In contrast, long-period giant planets are weakly irradiated and well-suited laboratories for testing relationships between mass, radius, and metal enrichment \citep{thorngren2016, teske2019}. Long-period giant planets that do not empirically show this radius anomaly allow independent investigations of interior structure, metallicity, and planet formation histories \citep{ginzburgchiang2020}. Additionally, long-period giant planets are testbeds for understanding the formation and evolution of hot Jupiters. They experience little to no tidal disruption and therefore their dynamical histories are better preserved as compared to their hot Jupiter counterparts.


In this paper, we report the planetary confirmation of \toib, a warm Jupiter on a 130-day eccentric orbit ($e\sim0.54$). Only one transit of \toib{} was recorded by the NASA \emph{TESS} mission to date. We conducted a ground-based follow-up campaign that eventually determined the period of the planet via both radial velocity orbit detection and subsequent photometric transit recovery. In Section\,\ref{sec:Observations}, we describe the photometric data from \tess and the ground-based LCO follow-up campaign, high-resolution speckle imaging, and CHIRON, FEROS, CORALIE, and PFS spectroscopic observations to obtain a planetary mass. Section\,\ref{sec:global} describes the global modeling of the system to derive system parameters. We then conclude with a discussion of the system in Section\,\ref{sec:discussion}.

\section{Observations} \label{sec:Observations}

\subsection{Photometric Observations} \label{subsec:photometry}
\subsubsection{TESS Photometry} \label{subsec:TESS}

\begin{figure*}
\begin{center}
\includegraphics[width=\textwidth]{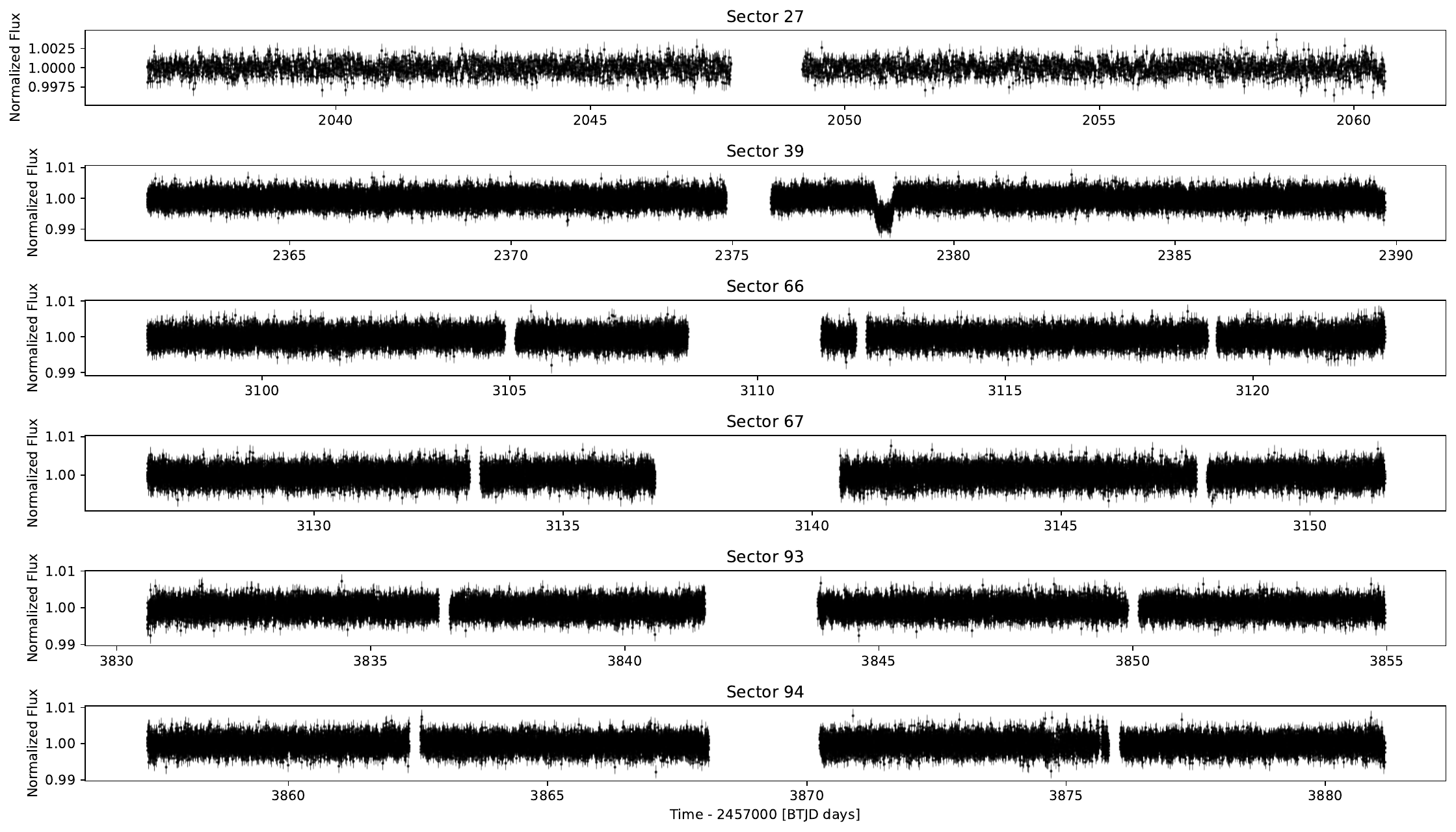}
\caption{Per-sector Normalized TESS PDCSAP light curves for TOI-6692. The target star was observed over six TESS sectors. The observing cadence was 600 seconds in Sector\,27 and 120\,seconds in Sectors 39, 66, 67, 93, and 94. Due to the long period, the planet had only a single TESS transit in Sector\,39.} 
\label{fig:raw_tess}
\end{center}
\end{figure*}

\begin{table*}
     \centering
     \caption{Confirmed Transiting Planets (6-20 Earth Radii) with $3\sigma$ Mass Measurements and Periods Longer than 100 days}
     \label{tab:longplanets}
     \begin{tabular}{lcccccccc}
     \hline\hline
     \textbf{Target} & \textbf{Period} &\textbf{Semi-Major Axis} & \textbf{Eccentricity} & \textbf{Radius} & \textbf{Mass} & \textbf{Multi-planet} & \textbf{Single Transit} & \textbf{Reference} \\
     & days & AU & & Jupiter & Jupiter & known & TESS &\\
     \hline
     \multicolumn{7}{l}{\textbf{Single Star Systems}}\\
    TOI-4465 b & 101.9  & 0.416 & 0.240 & 1.25 & 5.89 & No & Yes & \cite{TOI4465}\\
    TOI-199 b & 104.9  & 0.425 & 0.090 & 0.81 & 0.17 & Yes & Yes & \cite{toi199b}\\
    Kepler 539 b & 125.6  & 0.499 & 0.390 & 0.75 & 0.97 & Yes & No & \cite{Kepler539}\\
    \textbf{\toib} & \textbf{130.57}  & \textbf{0.512} & 0.\textbf{537} & \textbf{1.04} & \textbf{0.62} & \textbf{Possible} & Yes & \textbf{this work} \\
    KOI-3680 b & 141.2  & 0.534 & 0.496 & 0.99 & 1.93 & No & No & \cite{KOI3680}\\
    TOI-2010 b & 141.8  & 0.552 & 0.212 & 1.05 & 1.29 & No & Yes & \cite{mann2023}\\
    TIC 241249530 b & 165.8  & 0.641 & 0.941 & 1.19 & 4.98 & No & Yes & \cite{TIC241249530}\\
    Kepler 1514 b & 217.8  & 0.753 & 0.401 & 1.11 & 5.28 & Yes & No & \cite{kepler1514b}\\
    Kepler 111 c & 224.8  & 0.751 & 0.176 & 0.63 & 0.70 & Yes & No & \cite{Kepler111c-conf}\\
    TOI 4562 b & 225.1  & 0.768 & 0.760 & 1.12 & 2.30 & Yes & No & \cite{TOI4562b}\\
    TOI-2180 b & 260.8  & 0.828 & 0.368 & 1.01 & 2.76 & No & Yes & \cite{TOI2180}\\
    PH 2 b & 282.5  & 0.833 & 0.215 & 0.83 & 0.74 & No & No & \cite{Kepler111c-conf}\\
    Kepler 553 c & 328.2  & 0.898 & 0.346 & 1.03 & 6.70 & Yes & No & \cite{Kepler111c-conf}\\
    \multicolumn{7}{l}{\textbf{Known Multi Star Systems}}\\
    HD80606 b & 111.4  & 0.457 & 0.930 & 1.07 & 4.38 & No & No & \cite{HD80606}\\
    Kepler 35 b & 131.5  & 0.603 & 0.042 & 0.73 & 0.127 & No & No & \cite{Kepler35}\\    
    Kepler 16 b & 228.8  & 0.705 & 0.007 & 0.75 & 0.33 & No & No & \cite{Kepler16}\\
 \hline
     \end{tabular}
 \end{table*}

The Transiting Exoplanet Survey Satellite (\textit{TESS}; \citealt{ricker2015}) is performing an all-sky survey in search of transiting exoplanets around nearby bright host stars observing 24 $\times$ 96 degrees of the sky during sectors of approximately 27 days. \toi (\tic) was observed by \textit{TESS} in Sector\,27 at a cadence of 600\,seconds and in Sectors 39, 66, 67, 93, and 94 at a cadence of 120 seconds. Data were processed by the NASA Science Processing Operations Center pipeline (SPOC; \citealt{jenkins2016}) and the light curves were downloaded from the Mikulski Archive for Space Telescopes (MAST)\footnote{https://mast.stsci.edu/portal/Mashup/Clients/Mast/Portal.html} using the  \textit{Lightkurve} package \citep{lightkurve}. The Presearch Data Conditioning Simple Aperture Photometry (PDCSAP; \citealt{2012PASP..124..985S,2014PASP..126..100S,2012PASP..124.1000S}) light curves were used in our analysis and are plotted in Figure\,\ref{fig:raw_tess}. 

The SPOC pipelines did not identify this target as a TESS Object of Interest (TOI) because only a single transit was observed in the sectors. Citizen scientists from the Visual Survey Group (VSG; \citealt{VSG}) identified a single transit in Sector\,39 (2021\,May\,27 to Jun\,24) on 2021\, Aug\,07 during part of a focused survey for such long-period candidates which would not have typically triggered a standard pipeline detection. The candidate was also concurrently and independently discovered through manual vetting by the Planet Hunters TESS collaboration \citep{planethunters} as a Community TOI (cTOI) on 2021\,Oct\,14 and on 2023\,Oct\,05 it was promoted to a TOI \citep{Guerrero2021}. 


\subsubsection{Ground-based Photometric Monitoring Campaign} \label{subsec:groundphot}
\toib has a transit duration of $11.06 \pm 0.24$ hr, which is nearly impossible to observe from any single ground-based observatory.  Additionally, given that this planet was only observed via a single transit with TESS and the orbital period confirmed through radial velocities (RVs) as part of this work, the uncertainty in predicted future transit times was fairly large. Based on the ephemeris from the global fit (Section\,\ref{sec:global}) using the single TESS transit observation and the radial velocities, the $1\sigma$ transit uncertainty at the epoch of our ground-based follow-up was $-3.9$ days and $+4.6$ days. We alerted the TESS Follow-up Observing Program\footnote{https://tess.mit.edu/followup} Sub Group 1 \citep[TFOP;][]{TFOP} and started monitoring nightly with the Las Cumbres Observatory Global Telescope \citep[LCOGT;][]{Brown:2013} for 12 days surrounding the nominal predicted ephemeris from the RV orbital solution. 

LCOGT is a network of 1\,m robotic telescopes equipped with $4096\times4096$ SINISTRO cameras with an image scale of $0\farcs389$ per pixel, resulting in a $26\arcmin\times26\arcmin$ field-of-view. We observed 6 transit attempts at Cerro Tololo Inter-American Observatory in Chile (CTIO), 6 transit attempts with the South Africa Astronomical Observatory near Sutherland, South Africa (SAAO), and 2 transit attempts with the Siding Spring Observatory near Coonabarabran, Australia (SSO). A total of 14 observations were obtained over a 12 day period starting on UT 2025\,May\,13 (see Table\,\ref{tab:photometry}). We used the {\tt TESS Transit Finder}, a customized version of the {\tt Tapir} software package \citep{jensen2013}, to automatically schedule ground-based transit observations. All images were calibrated by the standard LCOGT {\tt BANZAI} pipeline \citep{McCully:2018}.  {\tt AstroImageJ} \citep[AIJ;][]{aij} was used to extract the differential aperture photometry. 

\begin{figure*}
 \includegraphics[width=\textwidth]{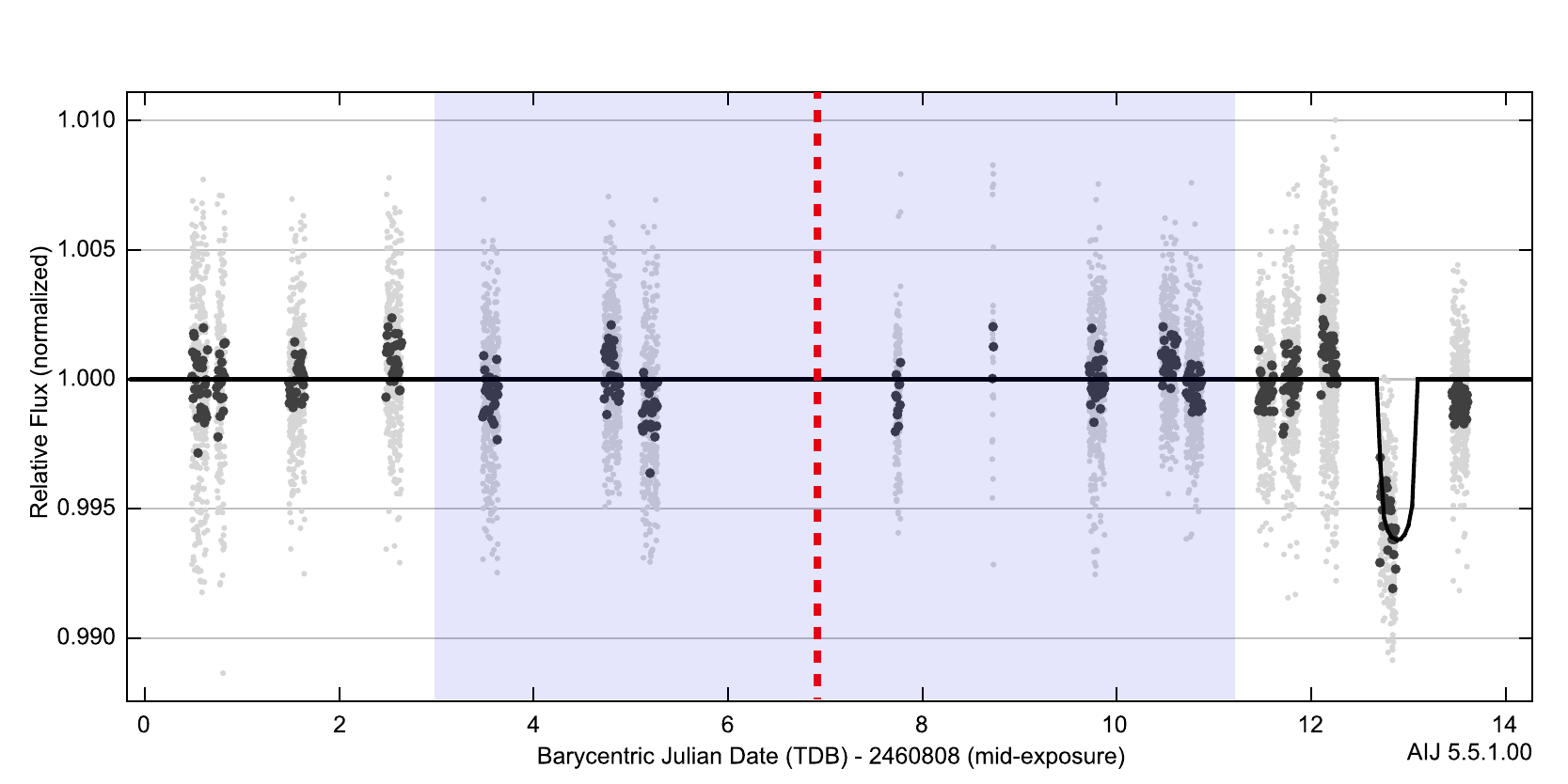} 
    \caption{Multi-night LCOGT 1.0\,m ground-based follow-up lightcurve. The differential lightcurve was extracted using a common set of 22 reference stars and was normalized over the full dataset. No lightcurve detrending was applied. Grey symbols show the unbinned data. Black symbols show the data in 8 minute bins. The vertical red dashed line shows the nominal mid-transit time based on the updated ephemeris from the global fit (Section\,\ref{sec:global}), which is based on the single TESS transit detection and all available radial velocities. The blue shaded region indicates the mid-transit time $1 \sigma$ uncertainty window. The apparent 6 ppt in-transit detection is just outside the $1 \sigma$ window. The bounds on orbital period based on the \tess single transit, RVs, and the apparent LCOGT in-transit detection is $P=131.125\pm 0.012$ days.}\label{fig:ground-phot}
 \end{figure*}

The time-series coverage of all observations is shown in Figure\,\ref{fig:ground-phot}. The $11.04$ hr transit duration is longer than the target is observable at any of the observatory locations. Unfortunately, several nights of bad weather occurred near the nominal transit center time prediction. An additional observation was obtained on 2025\,May\,19 UT with the TRAnsiting Planets and PlanetesImals Small Telescope in the South (TRAPPIST-South). TRAPPIST-South is a 60-cm robotic telescope located at La Silla Observatory in Chile \citep{gillon2011, jehin2011}. The telescope is equipped with an Andor iKon-L BEX2-DD deep-depletion 2K$\times$2K e2V CCD, resulting in a 20 $\times$ 20 arcminute FOV and a plate scale of 0.60" per pixel. Several images in the dataset were saturated and the observations were taken with a different CCD and filter than the others in this campaign. The TRAPPIST-South observation was excluded from the LCOGT final analysis in order to create a homogeneous set of observations, all taken in the same Sloan $i'$ filter, same telescope size, and same SINISTRO detectors. In addition, we used the same $7\farcs8$ radius photometric aperture and a common set of 22 reference stars to extract the combined differential lightcurve. 

The combined lightcurve was then normalized as a single lightcurve and no detrending was applied. The resulting lightcurve is shown in Figure\,\ref{fig:ground-phot}. There are 1-2 ppt offsets in the night-to-night, telescope-to-telescope, baselines, except for an apparent 5-6 ppt offset in the next-to-last observation. This offset could be due to the observation catching an in-transit segment of the expected 6 ppt deep transit. If this segment of data is indeed an in-transit detection, upper and lower bounds on the orbital period based on the \tess single transit, RVs, and the apparent LCOGT in-transit detection are $P=131.125\pm0.012$ days. The apparent event occurred just beyond the time window of the $1\sigma$ upper range, as determined from the global analysis in Section\,\ref{sec:global}. 

 \begin{table*}
     \centering
     \caption{LCO Ground-based Photometry Campaign}
     \label{tab:photometry}
     \begin{tabular}{lccccccc}
     \hline\hline
     \textbf{Observatory} & \textbf{Telescope size} &\textbf{Camera} & \textbf{Filter} & \textbf{Pixel Scale} & \textbf{UT Start Date} & \textbf{UT Time} & \textbf{RMS}  \\
     & meters & & & arseconds & yyyy-mm-dd & hh:mm:ss & ppt/10 min\\
     \hline
CTIO & 1.0  & SINISTRO & $i^{'}$ & 0.39 & 2025-05-13 & 05:43:08 - 07:52:12 & 0.69\\
SAAO & 1.0  & SINISTRO & $i^{'}$ & 0.39 & 2025-05-13 & 23:34:00 - 03:27:25 & 0.69\\
SAAO & 1.0  & SINISTRO & $i^{'}$ & 0.39 & 2025-05-14 & 23:29:59 - 03:23:37 & 0.66\\
SAAO & 1.0  & SINISTRO & $i^{'}$ & 0.39 & 2025-05-15 & 23:26:27 - 03:20:23 & 0.67\\
CTIO & 1.0  & SINISTRO & $i^{'}$ & 0.39 & 2025-05-17 & 05:27:30 - 09:21:17 & 0.69\\
SSO & 1.0  & SINISTRO & $i^{'}$ & 0.39 & 2025-05-17 & 14:46:19 - 18:39:47 & 0.69\\
CTIO & 1.0  & SINISTRO & $i^{'}$ & 0.39 & 2025-05-22 & 05:08:01 - 09:01:26 & 0.53\\
SAAO & 1.0  & SINISTRO & $i^{'}$ & 0.39 & 2025-05-22 & 22:58:36 - 02:52:17 & 0.51\\
CTIO & 1.0  & SINISTRO & $i^{'}$ & 0.39 & 2025-05-23 & 05:05:30 - 08:59:18 & 0.47\\
SAAO & 1.0  & SINISTRO & $i^{'}$ & 0.39 & 2025-05-23 & 22:54:50 - 02:48:21 & 0.50\\
CTIO & 1.0  & SINISTRO & $i^{'}$ & 0.39 & 2025-05-24 & 05:00:12 - 08:53:49 & 0.71\\
SSO & 1.0  & SINISTRO & $i^{'}$ & 0.39 & 2025-05-24 & 14:24:57 - 18:18:47 & 0.76\\
CTIO & 1.0  & SINISTRO & $i^{'}$ & 0.39 & 2025-05-25 & 04:56:18 - 08:49:56 & 0.67\\
SAAO & 1.0  & SINISTRO & $i^{'}$ & 0.39 & 2025-05-25 & 22:46:49 - 02:40:41 & 0.38\\
 \hline
     \end{tabular}
 \end{table*}

\subsubsection{High-resolution Imaging} \label{subsec:imaging}

Spatially close stellar companions to exoplanet host stars (bound or line-of-sight) can create a false-positive transit signal and will contribute ``Third-light” flux leading to an underestimated planetary radius \citep{ciardi2015} and incorrect derived parameters for both the host star and the planet if not accounted for \citep{furlan2017,furlan2020}. Thus, to search for close-in (bound) companions unresolved in TESS or other ground-based follow-up observations, we obtained high-resolution imaging speckle observations of \toi.

\begin{figure}
   \centering    \includegraphics[width=.9\linewidth,height=175pt]{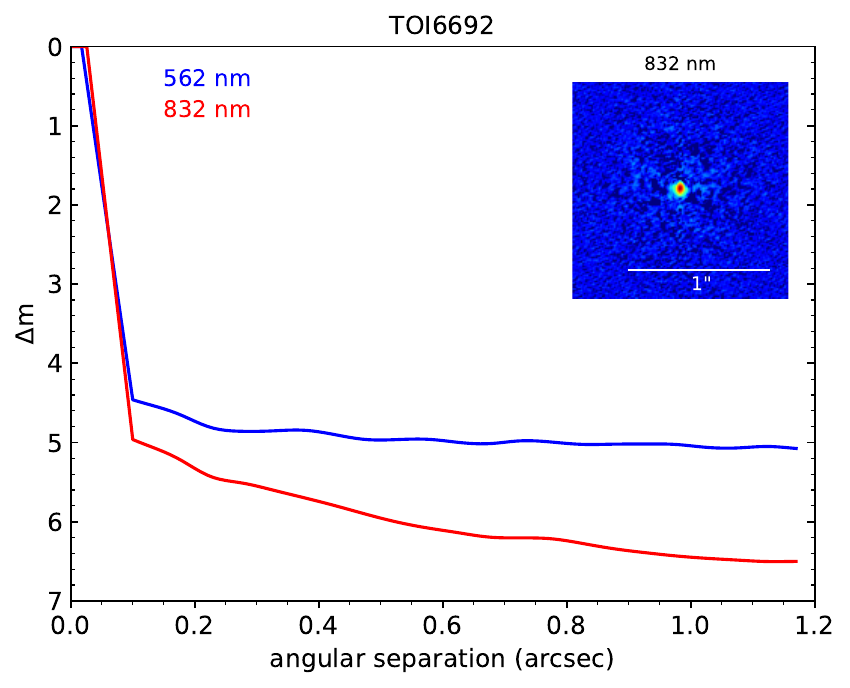} 
    \caption{ The figure shows $5\sigma$ magnitude contrast curves in both filters as a function of the angular separation out to 1.2 arcsec. The inset shows the reconstructed 832 nm image of \toi with a 1 arcsec scale bar. \toi was found to have no close companions from 0.1 to 1.2 arcsec to within the magnitude contrast levels achieved.}\label{fig:AO}
 \end{figure}

\toi was observed on 2025\,May\,10 UT using the Zorro speckle instrument on the Gemini South 8-m telescope \citep{scott2021}. Zorro provides simultaneous speckle imaging in two bands (562 nm and 832 nm) with output data products including a reconstructed image with robust contrast limits on companion detections (e.g., \cite{howell2025}). Four sets of 1000$\times$0.06 sec exposures were collected and subjected to Fourier analysis in our standard reduction pipeline (see \cite{howell2011}). Figure\,\ref{fig:AO} shows our final contrast curves and the 832 nm reconstructed speckle image. We find that \toi is a single star with no companion brighter than 5-6.5 magnitudes below that of the target star from 0\farcs1 out to 1\farcs2. At the distance of TOI-6692 (313\,pc) these angular limits correspond to spatial limits of 30 to 376\,au.


\subsection{Spectroscopic Observations} \label{subsec:spec}
A total of 89 spectra from four observatories were obtained between UT 2021\,Sep\,13 and 2025\,Jun\,17 to measure stellar parameters of the star and to measure a mass of the companion. Information about the facilities and observations is described in the following subsections.

 \begin{table}
     \centering
     \caption{PFS Radial Velocities}
     \label{tab:rvs}
     \begin{tabular}{rrc}
     \hline\hline
     \textbf{Time} & \textbf{Velocity} & \textbf{Uncertainty} \\
     \textbf{[\bjdtdb]} & \textbf{[\ms]} & \textbf{[\ms]}  \\
     \hline
2459717.91484   & 17.92   & 4.26  \\ 
2459796.68778   & 18.94   & 4.28  \\
2459834.59932   & -5.49   & 5.20  \\
2459834.60691   & -2.01   & 6.68  \\
2459853.55921   & 11.23   & 3.93  \\ 
2459855.53952   & 18.72   & 4.17  \\
2459889.52721   & 56.80   & 3.75  \\
2459893.53582   & 56.52   & 4.90  \\
2459895.54528   & 31.51   & 4.60  \\
2460065.89343   & 15.00   & 3.96  \\
2460065.90388   &  4.58   & 3.91  \\
2460067.89668   & -5.37   & 4.88  \\
2460067.90752   &   6.71  & 4.63  \\
2460072.89346   & -10.69  & 4.28  \\
2460072.90625   &   2.98  & 4.39  \\
2460073.90792   &  14.11  & 4.23  \\
2460073.91849   &   8.69  & 3.77  \\
2460074.89392   &  10.05  & 4.09  \\
2460074.90508   &  -6.84  & 3.81  \\
2460122.76347   &  -9.90  & 4.86  \\
2460122.77463   &  -4.70  & 4.74  \\
2460126.79969   &  -5.89  & 4.03  \\
2460126.81056   &  13.39  & 3.90  \\
2460153.74789   &  53.08  & 4.10  \\
2460180.64172   &  11.78  & 3.70  \\
2460210.67252   & -12.13  & 4.07  \\
2460459.84529   & -16.20  & 3.52  \\
2460461.85411   & -13.58  & 5.26  \\
2460463.83988   & -11.31  & 3.26  \\
2460464.85629   & -11.02  & 3.27  \\
2460485.79046   & -25.97  & 4.17  \\
2460488.84101   & -18.70  & 3.62  \\
2460490.77108   &  -4.65  & 3.35  \\
2460492.77987   &  -8.22  & 3.25  \\
2460509.75949   &  -7.42  & 3.27  \\
2460510.71692   &  -5.08  & 3.79  \\
2460516.71909   &   5.29  & 3.99  \\
2460517.72135   &   0.00  & 4.11  \\
2460536.67995   &  36.43  & 3.78  \\
2460544.67713   &  33.60  & 3.86  \\
2460596.54190   & -14.47  & 3.39  \\
2460805.90984   &  24.70  & 3.84  \\
2460810.89294   &  12.61  & 3.50  \\ 
2460834.84703   & -10.91  & 3.87  \\
2460843.82105   &  -5.51  & 4.30  \\
    \hline
     \end{tabular}
 \end{table}

 \subsubsection{CHIRON} \label{subsec:Chiron}
CHIRON is a fiber-fed echelle spectrograph on the 1.5m SMARTS telescope at Cerro Tololo Inter-American Observatory, Chile \citep{2013CHIRON}. CHIRON observes over the wavelength range $4100-8200$\,\AA\ and has a spectral resolving power of $R\sim 80,000$. Spectra were extracted as per the official CHIRON pipeline \citep{2021CHIRONreduction}. Thirty-one observations were obtained between September 2021 and September 2022. 

 \subsubsection{FEROS} \label{subsec:feros}
The Fiber-fed Extended Range Optical Spectrograph \citep[FEROS][]{feros} is a high resolution, temperature stabilized echelle spectrograph installed at the MPG-2.2m telescope, in the ESO La Silla Observatory, in Chile. FEROS has a resolving power of 48,000 and uses a second fiber to trace instrumental wavelength shifts. \textbf{A total of 6 observations} were obtained between 2023\,Aug and 2024\,Jun. All reduction and processing steps were performed using the \texttt{ceres} pipeline \citep{ceres}.

 \subsubsection{CORALIE} \label{subsec:coralie}
CORALIE \citep{1996BarraneCORALIE,2000CORALIE, 2010CORALIE} is a fiber-fed echelle spectrograph installed on the Swiss 1.2-m Leonhard Euler Telescope at the La Silla Observatory in Chile.  It has a spectral resolving power of $R\sim60,000$ and operates in the wavelength range 390-680\,nm. We observed \toi with the first fiber and we used the second fiber to observe the simultaneous Fabry–Pérot étalon for drift calibration purposes. A total of 7 spectra with an exposure time of 1800s were obtained between 2024\,Apr and Jul. The spectra were reduced with the standard data reduction pipeline. The radial velocity measurements were obtained by cross-correlating each spectrum with a G2-type stellar mask (e.g. \cite{2002pepeCORALIEccf}).

\subsubsection{PFS} \label{subsec:pfs}
The Planet Finder Spectrograph \citep[PFS;][]{crane2010PFS,crane2008PFS, crane2006} is a high-precision spectrograph on the 6.5m Magellan\,II Telescope at Las Campanas Observatory in Chile. A total of 45 spectra were obtained between 2022\,May and 2024\,Jun and the spectra were reduced using the methods outlined in \cite{1996butler}. The spectra were collected using a 0.3\,arcsecond wide slit, $3\times3$ binning, and 1200\,second exposures. The spectral resolving power in this configuration is $R\sim115,000$.

\subsection{SED analysis} \label{sec:sed}
\begin{figure}
     \centering    \includegraphics[width=0.46\textwidth]{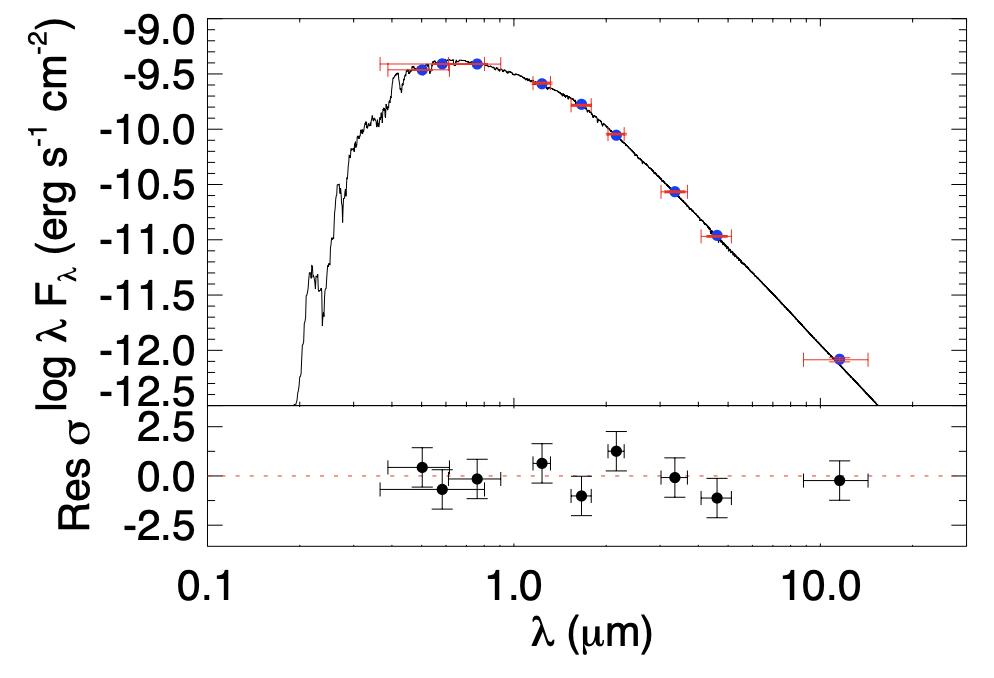}
     \caption{Spectral energy distribution of the target star \toi. Magnitudes from \emph{Gaia} $G$, $B_p$, $R_p$, 2MASS $J$, $H$, $Ks$, and WISE $W1$, $W2$, $W3$, and $W4$ are included in the global modeling of the system and are shown in dark blue. }\label{fig:sed}
 \end{figure}

All available broadband photometry, including  \emph{Gaia} DR3 $G$, $Bp$, $Rp$ \citep{2022arXiv220800211G}, 2MASS $J$, $H$, $K$ \citep{2006AJ....131.1163S}, and WISE $W_1$, $W_2$, $W_3$ bands \citep{WISE}, as well as \emph{Gaia} DR3 parallax were used to construct the spectral energy distribution (SED) of \toi. The spectral energy distribution was modeled simultaneously with the photometric and spectroscopic observations of the system which is discussed in the next Section. The best-fit model is shown in Figure\,\ref{fig:sed} and the catalog parameters are reported in Table\,\ref{tab:star}. 

 \begin{table}
     \caption{Literature values for \toi.
     \tablewidth{\columnwidth}
     \label{tab:star}}
     \centering
     \begin{tabular}{lll}
     \hline\hline
 \textbf{Stellar Parameters} & \textbf{Value} & \textbf{Source}\\
 \hline
 \multicolumn{3}{l}{\textbf{Catalog Information}}\\
 TIC ID & \tic &   TOI Catalog \\
 TOI ID & \toi &  TOI Catalog \\
 \textit{Gaia} DR3 ID & 6349145498210403840 & GAIA DR3\\
 \textit{2MASS} ID & J20504634-8116202 & 2MASS\\
 TYC ID & 9473-00833-1 & Tycho\\
 \multicolumn{3}{l}{\textbf{Coordinates and Proper Motion}} \\
 Right Ascension (h:m:s)  & 20:50:46.57 (J2000) &  GAIA DR3 \\
 Declination (d:m:s)    & -81:16:20.63  (J2000) &  GAIA DR3 \\
 Parallax (mas)   & $\parallax$ &  GAIA DR3 \\
 $\mu$\textsubscript{R.A.} (mas yr\textsuperscript{-1}) & $\pmRA$ & GAIA DR3 \\
 $\mu$\textsubscript{Dec.} (mas yr\textsuperscript{-1}) & $\pmDEC$ &  GAIA DR3 \\
 \multicolumn{3}{l}{\textbf{Magnitudes}} \\
 $TESS$ (mag)         & $11.116 \pm 0.006$ &  TOI Catalog \\
 $G$ (mag)         & 11.578162~$\pm$~0.002762 & GAIA DR3 \\
 $B_{p}$ (mag)         & 11.932896~$\pm$~0.002829 & GAIA DR3 \\
 $R_{p}$ (mag)         & 11.050061~$\pm$~0.003797 & GAIA DR3 \\
 $B$ (mag)      & 12.305~$\pm$~0.209 & Tycho \\
 $V$ (mag)         & 11.434~$\pm$~0.016 & Tycho \\
 $J$ (mag)         & 10.432~$\pm$~0.023 & 2MASS \\
 $H$ (mag)         & 10.136~$\pm$~0.022 & 2MASS \\
 $K$ (mag)        & 10.034~$\pm$~0.019 & 2MASS\\
 $WISE_{3.4\mu}$ (mag)     & 10.029~$\pm$~0.023 & WISE \\
 $WISE_{4.6\mu}$ (mag)     & 10.053~$\pm$~0.020 & WISE \\
 $WISE_{12\mu}$ (mag)     & 10.010~$\pm$~0.046 & WISE \\
 \hline
 \end{tabular}
 \begin{flushleft}
 \footnotesize{\textbf{Note:} TESS TOI Primary Mission Catalog; \citep{Guerrero2021}, Tycho; \citep{tycho}; GAIA DR3; \citep{Gaiadr3}, 2MASS; \citep{2MASS}, WISE; \citep{allwise}}
 \end{flushleft}
 \end{table}

\section{Global Modeling} \label{sec:global}
We performed a joint analysis using $\texttt{EXOFASTv2}$\xspace \citep{eastman2013, exofastv2} to determine the stellar and planetary parameters of the system. We incorporated the TESS photometry, all available radial velocities from PFS, CHIRON, FEROS, and CORALIE, and catalog observations to simultaneously fit the system. $\texttt{EXOFASTv2}$\xspace is an open-source code written in \texttt{IDL} that uses Markov Chain Monte Carlo (MCMC) to derive stellar and planetary parameters. Gaussian priors were adopted for parallax obtained from GAIA Data Release 3 \citep[GAIA DR3;][]{Gaiadr3} which was corrected by \cite{lindegren2021}. The starting values for the transit epoch, $T_{c}$, and the period, $P$, were set from the orbital fit of the radial velocities and the initial starting values on the stellar mass, $M_{*}$, and radius, $R_{*}$, were obtained from the TESS mission input catalog \citep{2018stassun}. We also set a Gaussian prior on metallicity derived from a \texttt{SpecMatch-Synth} stellar parameter analysis \citep{SpecMatchSynth_Petigura2015} of the iodine-free PFS template spectrum. In brief, \texttt{SpecMatch-Synth} matches the observed spectrum with the \citet{Kurucz1993} grid of synthetic stellar spectra. The code interpolates between the nearest grid spectra by taking linear combinations to obtain the final spectroscopic parameters. We measured a slightly sub-solar metallicity for \toi of $\feh = -0.10 \pm 0.06$~dex from this analysis. 

\begin{figure}
   \centering    \includegraphics[width=.9\linewidth,height=175pt]{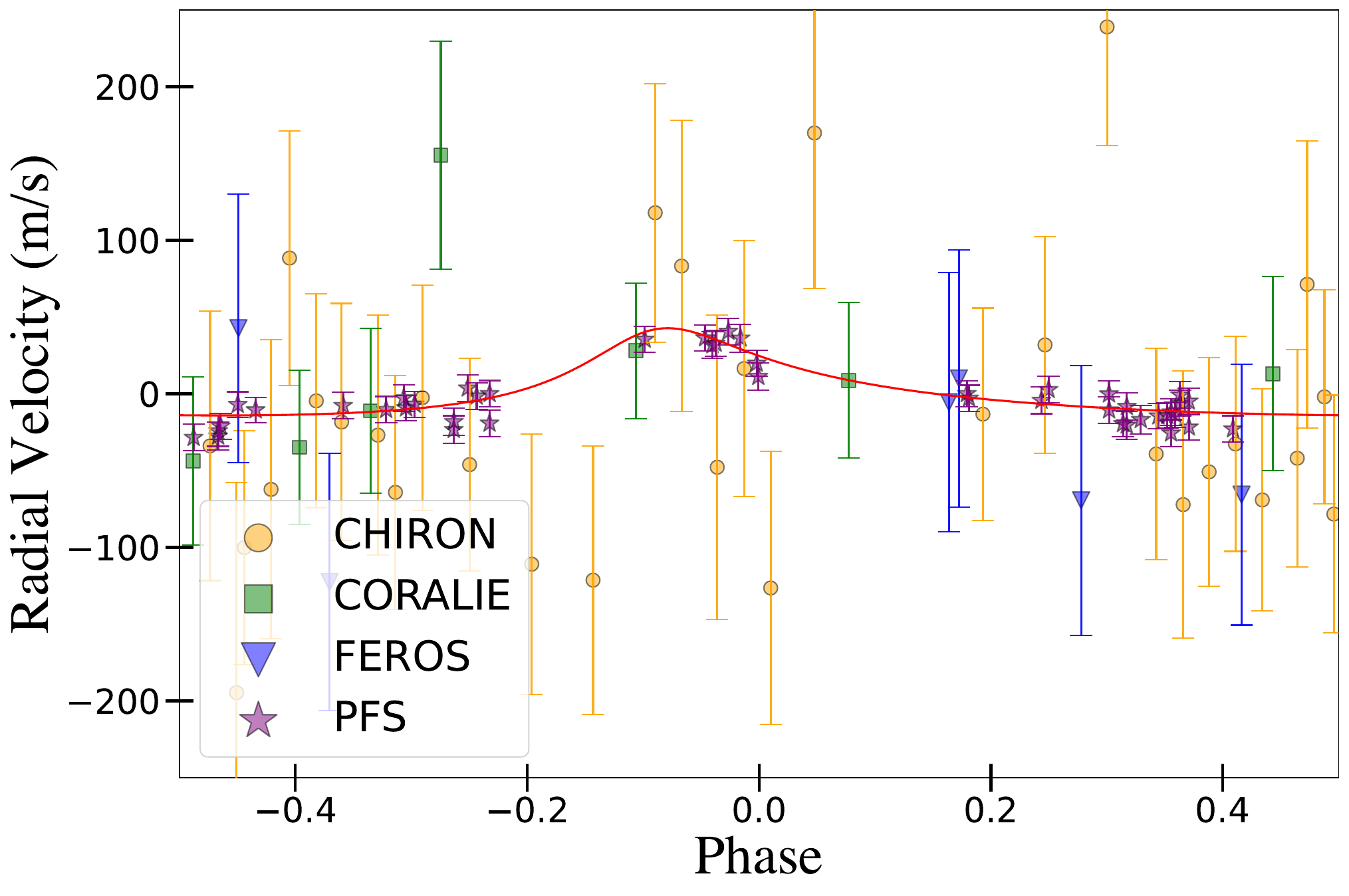} 
    \caption{Radial velocity observations phase folded to the ephemeris. The red line shows the $\texttt{EXOFASTv2}$ best model fit to the RVs. Section\,\ref{subsec:spec} details the radial velocities observations.}\label{fig:ALLrvs}
 \end{figure}

\begin{figure}
   \centering    \includegraphics[width=.9\linewidth,height=300pt]{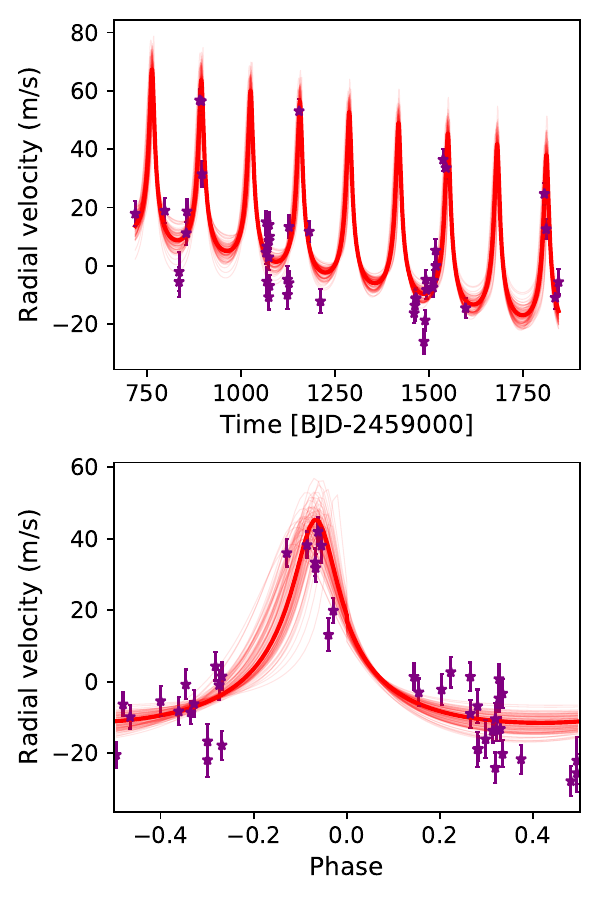} 
    \caption{\textbf{Top:} A linear slope showing the PFS radial velocities over time. The trend is detected at $-0.0280^{+0.0046}_{-0.0045}\,\mathrm{m\,s}^{-1}\mathrm{day}^{-1}$. Longer term monitoring is necessary to confirm the orbital period of the outer companion. \textbf{Bottom:} PFS radial velocity observations phase folded to the ephemeris for better clarity. The solid bold red line is the best fit model from $\texttt{EXOFASTv2}$ using RVs from all four facilities. The finer red lines show a small selection of draws from the converged posterior. Section\,\ref{subsec:spec} details the radial velocities observations. }\label{fig:PFSrvs}
 \end{figure}

We note that the CHIRON, CORALIE, and FEROS RVs have much lower precision than the PFS RVs as shown in Figure \ref{fig:ALLrvs}. We opted to run comparison fits to ensure that the lower-precision RVs were not degrading the result. In both fits, we included the TESS photometry and all available catalog observations as noted above. We found that in both fits, the stellar and planetary parameters agreed within 1$\sigma$. Therefore, we concluded that the lower-precision RVs did not degrade the fit and we included them in the fit that we adopted as our final results. Those results are listed in Table\,\ref{tab:bestfit} which also lists the prior values used. Each fit also allowed for a long-term linear trend of $-0.0280^{+0.0046}_{-0.0045}$ m/s/day as shown in Figure \ref{fig:PFSrvs}. This indicates that there may be an unresolved long-period companion in the system. We also ran a comparison two-planet model to check if the second companion's orbital period could be constrained. $\Delta\rm{BIC}$ preferred the single-planet plus RV trend model. We detect no additional periodic transits in the photometry nor do we detect any companions in the high-resolution imaging. Companion threshold detection limits are described in more detail in the Discussion. 

We did not include the ground-based photometric follow-up data in the global fits because the detection is an in-transit event and the data have gaps in time. A global fit that included this data would be very time intensive and not well constrained due to the lack of ingress or egress detected. We instead describe our photometric analysis in Section \ref{subsec:groundphot} and provide an alternative updated period from this analysis for reference.

\begin{table*}
\centering
\caption{Best-fit Stellar and Planetary Properties for \toi}
\label{tab:bestfit}
\begin{tabular}{lllr}
\hline\hline
Parameters & Description (Units) & Prior Values & Best Fit \\
\hline
\multicolumn{4}{l}{\textbf{Stellar Parameters:}} \\
$M_{\star}$   & Stellar Mass ($M_\odot$)       & Inferred                   & $\mstar$ \\
$R_{\star}$   & Stellar Radius ($R_{\odot}$)   & Inferred                   & $\rstar$ \\
$L_{\star}$   & Stellar Luminosity ($L_{\odot}$) & Inferred                 & $\lstar$ \\
$T_{\rm eff}$ & Effective Temperature (K)      & Inferred                   & $\tefffit$ \\
$\log g$      & Surface Gravity (cgs)          & Inferred                   & $\loggfit$ \\
$\mathrm{[m/H]}$ & Metallicity (dex)          & $\mathcal{G}(-0.1,0.06)$   & $\metfit$ \\
Parallax      & Parallax (mas)                 & $\mathcal{G}(3.220,0.012)$ & $3.220\pm0.012$ \\
Age           & Age (Gyr)                      & Inferred                   & $\age$ \\
Distance      & Distance (pc)                  & Inferred                   & $\dist$ \\
$A_V$         & V-band extinction (mag)        & $\mathcal{U}(0.00, 0.48)$  & $0.29^{+0.11}_{-0.13}$ \\
\hline
\multicolumn{4}{l}{\textbf{Planetary Parameters:}} \\
$P$           & Orbital Period$^{(1)}$ (days)          & Inferred                  & $\per$ \\
$P$           & Updated Orbital Period$^{(2)}$ (days)  & Inferred                  & $131.125\pm 0.012$ \\
$T_{0}$       & Epoch (BJD)                    & $\mathcal{U}(2459378.4,2459378.5)$ & $2459378.4174\pm0.0022$ \\ 
$M_{p}$       & Planet Mass ($\mj$)            & Inferred                   & $\plmass$ \\
$R_{p}$       & Planet Radius ($\rj$)          & Inferred                   & $\plrad$ \\
$a/R_{\star}$ & Semi-major axis to star radius ratio & Inferred             & $79.5^{+4.2}_{-3.9}$ \\
$a$           & Semi-major axis (AU)           & Inferred                   & $\semimaj$ \\
$e$           & Eccentricity                   & Inferred                   & $\ecc$ \\
$\omega_*$    & Arg of periastron (deg)        & Inferred                   & $-16.6^{+7.9}_{-7.7}$ \\
$T_{\rm eq}$  & Equilibrium temp$^{(3)}$ (K)   & Inferred                   & $467.0^{+9.3}_{-9.5}$ \\
$T_{14}$      & Transit duration (hours)       & Inferred                   & $11.06 \pm 0.24$ \\
$K$           & RV semi-amplitude (m/s)        & Inferred                   & $28.5^{+3.2}_{-2.7}$ \\
$\dot{\gamma}$& RV slope$^{(4)}$ (m/s/day)     & Inferred                   & $-0.0280^{+0.0046}_{-0.0045}$ \\
$i$           & Orbital inclination (deg)      & Inferred                   & $89.483^{+0.083}_{-0.074}$ \\
$b$           & Impact parameter               & Inferred                   & $0.618^{+0.072}_{-0.120}$ \\
\hline
\end{tabular}

\vspace{0.5ex}
{\footnotesize
(1) Derived from global analysis (Section\,\ref{sec:global}).\\
(2) Updated based on the ground-based lightcurve detection (Section\,\ref{subsec:groundphot}).\\
(3) Assumes no albedo and perfect redistribution.\\
(4) Reference epoch = 2460157.211885.\\
}
\end{table*}

 \section{Discussion}
 \label{sec:discussion}
\begin{figure}
    \centering
    \includegraphics[width=0.95\linewidth]{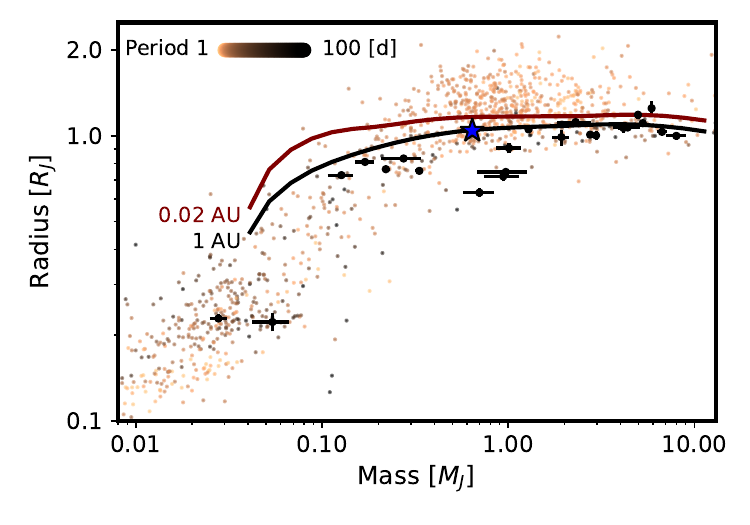} 
    \caption{Planet mass versus planet radius for all planets with $3\sigma$ mass and radii measurements. Planets with an orbital period $\geq 100$ days are plotted and shown in black with their respective error bars. Data were obtained on 2025\,Aug\,04 from the NASA Exoplanet Archive. \toib is shown as a blue star with black edges and the solid lines show the $10$\me\, core \citep{fortney2007} giant planet thermal and atmospheric models at 1\,AU in black and at 0.02\,AU in red.}\label{fig:mass-radius}
 \end{figure}

\begin{figure}
    \centering
    \includegraphics[width=0.95\linewidth]{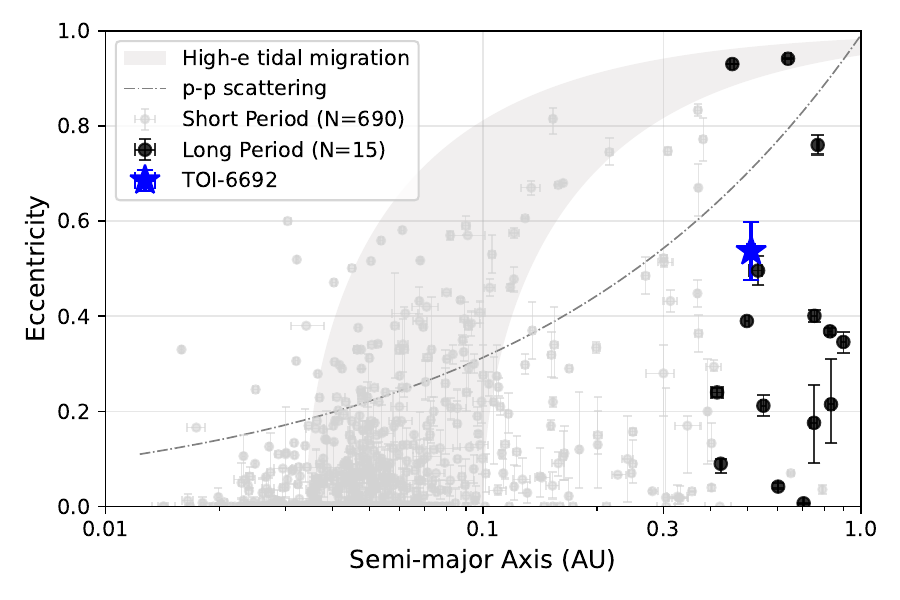}
    \caption{All transiting exoplanets in the range of 6-20 Earth radii. Planets with periods less than 100 days are plotted in light gray and the 16 planets with periods longer than 100 days are plotted in black. \toib is plotted as a blue star. The gray region illustrates planets that are likely undergoing high-eccentricity tidal migration. The upper and lower limits of the track are set by the Roche limit and the tidal circularization timescale \citep{toi3362}.  The dotted-dashed line presents the theoretical upper limit of eccentricities as a result of planet-planet scattering, assuming a planet with a mass of 0.5 $M_{\rm Jup}$ and a radius of 2 $R_{\rm Jup}$, for illustrative purpose \citep{Petrovich:2014}. Data were obtained on 2025\,Aug\,04 from the NASA Exoplanet Archive.}\label{fig:ecc}
 \end{figure}

\toib is a long-period Jupiter-mass planet in an eccentric ($e\sim0.54$) orbit. Only 15 giant planets have masses and radii measured (at $>3\sigma$ significance) with periods longer than 100 days (Table\,\ref{tab:longplanets}). Figures~\ref{fig:mass-radius} and ~\ref{fig:ecc} place \toib{} into context in the mass, radius, period and eccentricity space.

\toib is part of a dynamically complex, long-period Jupiter-like system. The planet has an eccentric orbit with a period of 130 days and an eccentricity of $\ecc$. PFS radial velocities detect a $\sim 20\,\mathrm{m\,s}^{-1}$ long term linear trend over a baseline of $\sim 800$ days. We combine the radial velocities and the limiting thresholds from speckle imaging to determine constraints on the mass and separation of this exterior companion. We simulate the Keplerian orbit of the exterior companion, and model its $I$ magnitude via the MIST isochrones \citep{dotter2016}. We adopt a distance of $310.5$ pc in these calculations as per the \emph{Gaia} parallax. The modeled Keplerian orbit is matched against the PFS velocity residuals after the removal of \toib{}. The luminosity of the companion and its semi-major axis were checked against the speckle imaging threshold in Section~\ref{subsec:imaging}. 

\begin{figure}
    \centering
    \includegraphics[width=0.95\linewidth]{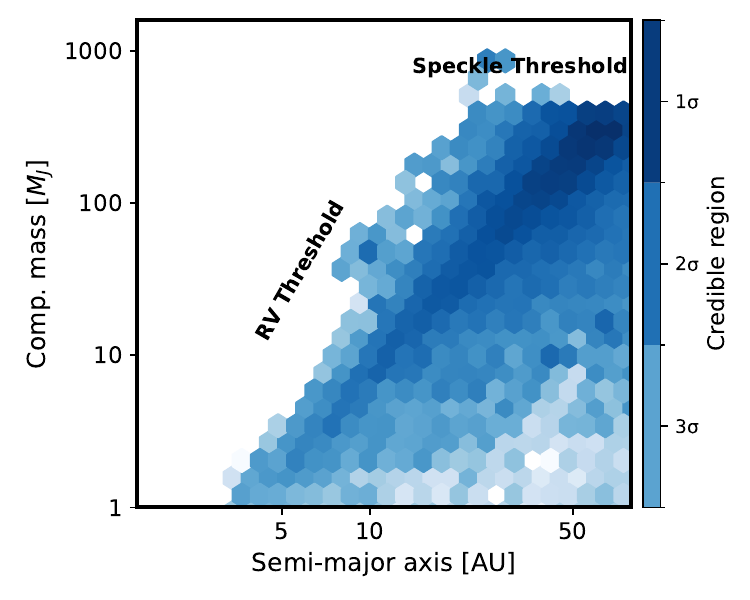}
    \caption{Limits on the properties of the exterior companion that we can place via PFS radial velocities and speckle imaging observations. The exterior companion is limited to $400\,M_J$ in mass from the speckle observations at $\sim 20$\,AU, and by radial velocities for interior orbital solutions. }
    \label{fig:companionlimits}
\end{figure}

Figure~\ref{fig:companionlimits} shows the constraints that we can place on this exterior companion. We place an upper mass limit of the companion to $\sim 400\,M_J$, from the speckle imaging observations at orbital distances $>20\,\mathrm{AU}$. Interior to this, the constraints are dominated by the radial velocities. 

The eccentricity of \toib{} is similar to other recent long-period giant planet discoveries from \emph{TESS} and \emph{Kepler} \citep{mann2023,Hebard2019}. As per Figure~\ref{fig:ecc}, at such large orbital separations, the eccentricity is not shaped by tidal circularization, and the planet is not expected to follow a high-eccentricity inward-migration track. In fact, amongst the transiting long-period (P $>100$ days) planets only HD80606b \citep{2001A&A...375L..27N,2009ApJ...703.2091W} and TIC241249530b \citep{2024Natur.632...50G} follow the classical high eccentricity migration track. A significantly larger population of planets has mild eccentricities $<0.8$. \citet{2008FordRasio} demonstrates that planet-planet scattering processes yield eccentricities no larger than $e\sim0.8$, and \toib{} is most consistent with migration via this process \citep[e.g.][]{1996Sci...274..954R, 2008Chatterjee,2008nagasawa}.

Figure~\ref{fig:mass-radius} shows \toib{} on the mass-radius distribution of giant planets, highlighting planets with periods $>100$\,days. These planets receive mild irradiation compared to conventional hot-Jupiters, and are excellent tests of traditional mass-radius relationships for giant planets. The mass and radius of \toib{} is in excellent agreement with model predictions from \citet{fortney2007}. The role tidal heating plays in retaining internal heat and an `inflated' radius of some planets remains unclear, but atmospheric studies have shown that some low density giant planets exhibit methane quenching in their atmospheres, suggesting significant interior heating \citep[e.g.][]{2024Natur.630..831S}. At transit, \toib{} is at a separation of 0.32\,AU, with an equilibrium temperature of $\sim 600$\,K.  At this temperature, methane should be a significant absorber at equilibrium. \toib{} has a transmission spectroscopic metric (TSM) of 23 \citep{2018PASP..130k4401K}. Among planets with similar irradiation, only four other giant planets have higher TSMs. These planets may serve as excellent laboratories for understanding the chemical composition of cold-but-accessible giant planets, to serve as comparison against similar planets closer in that may exhibit disequilibrium due to added internal heating or photochemistry.

\toib demonstrates the challenges associated with confirming long-period, single-transit planet candidates from TESS. While some constraints can be placed from single transits on the orbital period of systems based on the transit duration, the uncertainties are very large. It requires a significant investment of telescope time to confirm the true period. In particular, high-precision radial velocity and photometric follow-up resources are limited. In the case of \toib, once we confirmed the period and eccentricity with the radial velocity follow-up, we were then able to utilize the strength of the TFOP SG1 photometric resources and, in particular, the LCO global network of 1-m telescopes to observe the second transit and further refine the period. Similar efforts confirmed the transit of the $P = 542$ days planet HIP 41378 f \citep{2025arXiv250620907G}, the transit of TOI-2010b via space-based small-sat observations \citep{mann2023}, and the NGTS recovery of TIC238855958b \citep{2020MNRAS.491.1548G}.

\section{acknowledgments}

We respectfully acknowledge the traditional custodians of the lands on which we conducted this research and throughout Australia. We recognize their continued cultural and spiritual connection to the land, waterways, cosmos and community. We pay our deepest respects to all Elders, present and emerging people of the Giabal, Jarowair and Kambuwal nations, upon whose lands this research was conducted.
GZ thanks the support of the ARC Future Fellowship program FT230100517.
Funding for the TESS mission is provided by NASA's Science Mission Directorate. We acknowledge the use of public TESS data from pipelines at the TESS Science Office and at the TESS Science Processing Operations Center. This research has made use of the Exoplanet Follow-up Observing Program website, which is operated by the California Institute of Technology, under contract with the National Aeronautics and Space Administration under the Exoplanet Exploration Program. Resources supporting this work were provided by the NASA High-End Computing (HEC) Program through the NASA Advanced Supercomputing (NAS) Division at Ames Research Center for the production of the SPOC data products. This paper includes data collected by the TESS mission that are publicly available from the Mikulski Archive for Space Telescopes (MAST). This work has made use of data from the European Space Agency (ESA) mission
{\it Gaia} (\url{https://www.cosmos.esa.int/gaia}), processed by the {\it Gaia}
Data Processing and Analysis Consortium (DPAC,
\url{https://www.cosmos.esa.int/web/gaia/dpac/consortium}). Funding for the DPAC has been provided by national institutions, in particular the institutions participating in the {\it Gaia} Multilateral Agreement. The Flatiron Institute is a division of the Simons Foundation. This work makes use of observations from the LCOGT network. Part of the LCOGT telescope time was granted by NOIRLab through the Mid-Scale Innovations Program (MSIP). MSIP is funded by NSF. 
Some of the observations in this paper made use of the High-Resolution Imaging instrument Zorro and were obtained under Gemini Proposal Number: GS-2025A-DD-101. Zorro was funded by the NASA Exoplanet Exploration Program and built at the NASA Ames Research Center by Steve B. Howell, Nic Scott, Elliott P. Horch, and Emmett Quigley. Zorro was mounted on the Gemini South telescope of the international Gemini Observatory, a program of NSF’s OIR Lab, which is managed by the Association of Universities for Research in Astronomy (AURA) under a cooperative agreement with the National Science Foundation. on behalf of the Gemini partnership: the National Science Foundation (United States), National Research Council (Canada), Agencia Nacional de Investigación y Desarrollo (Chile), Ministerio de Ciencia, Tecnología e Innovación (Argentina), Ministério da Ciência, Tecnologia, Inovações e Comunicações (Brazil), and Korea Astronomy and Space Science Institute (Republic of Korea).
Funding for KB was provided by the European Union (ERC AdG SUBSTELLAR, GA 101054354). MG and EJ are F.R.S-FNRS Senior Research Directors. 
TRAPPIST is funded by the Belgian Fund for Scientific Research (Fond National de la Recherche Scientifique, FNRS) under the grant FRFC 2.5.594.09.F, with the participation of the Swiss National Science Fundation (SNF).
R.B. acknowledges support from FONDECYT Project 1241963 and from ANID -- Millennium  Science  Initiative -- ICN12\_009.
MK acknowledges the support of the Natural Sciences and Engineering Research Council of Canada (NSERC), RGPIN-2024-06452. Cette recherche a été financée par le Conseil de recherches en sciences naturelles et en génie du Canada (CRSNG), RGPIN-2024-06452.
T.D. acknowledges support from the McDonnell Center for the Space Sciences at Washington University in St. Louis.
We thank Adriana Kuehnel who contributed to some of the PFS observations and who built and maintains the PFS website used for decision making of observations.
\textbf{A.J.\ acknowledges support from Fondecyt project 1251439.}

\vspace{5mm}
\facilities{MAST(TESS), CHIRON, FEROS, CORALIE, PFS, LCOGT, ExoFOP}

\software{AstroImageJ \citep{aij}, Lightkurve \citep{lightkurve}, Tapir \citep{jensen2013}, Exofastv2 \citep{exofastv2}}

\bibliography{ref}
\bibliographystyle{aasjournal}

\end{document}